\pdfoutput=1
\documentclass[usenatbib]{mn2e}
\usepackage{apjfonts}
\usepackage{amssymb}
\usepackage{amsmath}
\usepackage{ctable}
\usepackage{fixltx2e} 

\newcommand{\be}{\begin{equation}}
\newcommand{\ee}{\end{equation}}

\newcommand{\cm}{{\rm cm}}
\newcommand{\etal}{et al.}

\newcommand{\msun}{M_{\sun}}

\newcommand{\paperone}{Paper {\small I}}
\newcommand{\papertwo}{Paper {\small II}}

\newcommand\plotonesize[2]
 {\centering \leavevmode \includegraphics[width={#2\columnwidth}]{#1}}
\newcommand{\plotsidesize}[2]
 {\centering \leavevmode \includegraphics[width={#2\textwidth}]{#1}}
\newcommand{\acknowledgments}{\begin{small}\section*{Acknowledgments}\end{small}}
\newcommand\altaffilmark[1]{$^{#1}$}
\newcommand\altaffiltext[1]{$^{#1}$}
\title[Dense Gas \&\ Stellar Feedback]{Dense Molecular Gas: A Sensitive Probe of Stellar Feedback Models\vspace{-0.5cm}}

\author[Hopkins \etal]{
\parbox[t]{\textwidth}{ 
Philip F.~Hopkins\thanks{E-mail:phopkins@astro.berkeley.edu}\altaffilmark{1},
Desika Narayanan\altaffilmark{2}, 
Norman Murray\altaffilmark{3,4}, \&
Eliot Quataert\altaffilmark{1}
}
\vspace*{6pt} \\
\altaffiltext{1}{Department of Astronomy and Theoretical Astrophysics Center, University of California Berkeley, Berkeley, CA 94720} \\
\altaffiltext{2}{Steward Observatory, University of Arizona, 933 
N Cherry Ave, Tucson, Az, 85721} \\
\altaffiltext{3}{Canadian Institute for Theoretical Astrophysics, 
60 St.\ George Street, University of Toronto, ON M5S 3H8, Canada} \\
\altaffiltext{4}{Canada Research Chair in Astrophysics} 
\vspace{-0.5cm}}

\date{Submitted to MNRAS, July, 2012\vspace{-0.6cm}}
\begin{document}
\maketitle
\label{firstpage}

\begin{abstract}

We show that the mass fraction of GMC gas ($n\gtrsim100\,{\rm cm^{-3}}$) in dense ($n\gg10^{4}\,{\rm cm^{-3}}$) star-forming clumps, observable in dense molecular tracers ($L_{\rm HCN}/L_{\rm CO(1-0)}$), is a sensitive probe of the strength and mechanism(s) of stellar feedback, as well as the star formation efficiencies in the most dense gas. Using high-resolution galaxy-scale simulations with pc-scale resolution and explicit models for feedback from radiation pressure, photoionization heating, stellar winds, and supernovae (SNe), we make predictions for the dense molecular gas tracers as a function of GMC and galaxy properties and the efficiency of stellar feedback/star formation. In models with weak/no feedback, much of the mass in GMCs collapses into dense sub-units, predicting $L_{\rm HCN}/L_{\rm CO(1-0)}$ ratios order-of-magnitude larger than observed.  By contrast, models with feedback properties taken directly from stellar evolution calculations predict dense gas tracers in good agreement with observations. Changing the strength or timing of SNe tends to move systems along, rather than off, the $L_{\rm HCN}-L_{\rm CO}$ relation (because SNe  heat lower-density material, not the high-density gas). Changing the strength of radiation pressure (which acts efficiently in the  highest density gas), however, has a much stronger effect on $L_{\rm HCN}$ than on $L_{\rm CO}$. 
We show that degeneracies between the strength of feedback, and efficiency of star formation on small scales, can be broken by the combination of dense gas, intermediate-density gas, and total SFR tracers, and favor models where the galaxy-integrated star formation efficiency in dense gas is low. We also predict that the fraction of dense gas ($L_{\rm HCN}/L_{\rm CO(1-0)}$) increases with increasing GMC surface density; this drives a trend in $L_{\rm HCN}/L_{\rm CO(1-0)}$ with SFR and luminosity which has tentatively been observed. Our results make specific predictions for enhancements in the dense gas tracers in unusually dense environments such as ULIRGs and galactic nuclei (including the galactic center).

\end{abstract}

\begin{keywords}
star formation: general --- galaxies: formation --- galaxies: evolution --- 
galaxies: active --- cosmology: theory
\vspace{-1.0cm}
\end{keywords}

\vspace{-1.1cm}
\section{Introduction}
\label{sec:intro}

Feedback from massive stars is critical to the evolution of galaxies, the properties and structure of the ISM, and the formation of stars and star clusters in giant molecular clouds. 
The Kennicutt-Schmidt law for star formation in galaxies implies a gas consumption time of $\sim50$ dynamical times \citep{kennicutt98}.
Moreover, the total fraction of the gas turned into stars in GMCs over their lifetime is only a few percent  \citep{zuckerman:1974.gmc.constraints,williams:1997.gmc.prop,evans:1999.sf.gmc.review,evans:2009.sf.efficiencies.lifetimes}.   

In an instantaneous sense, the low star formation rate in GMCs is closely related to the fact that most of the gas within GMCs is at relatively low densities $n\sim 10-100\cm^{-3}$, i.e., {\em not} in star-forming cores that have densities $\gtrsim 10^{4}\,{\rm cm^{-3}}$ \citep[e.g.][and references therein]{williams:1997.gmc.prop,evans:1999.sf.gmc.review}. 
Tracers of high-density gas such as the HCN transition (critical density $n\gtrsim 10^{4}\,{\rm cm^{-3}}$) have shown that it is the gas at these high densities that is actually forming stars -- what determines the SFR in GMCs is the amount of mass at these densities \citep[e.g.][]{shirley:2003.core.mapping,wu:2005.hcn.properties,lada:2010.sf.traces.dense.gas}. In typical GMCs, this is $\sim 10\%$ of the mass, but star formation is relatively rapid in the clumps, giving the canonical SFR of a few percent of the GMC mass per dynamical time. 

Furthermore, observations have suggested that in high-density systems such as local merger-induced starbursts -- which are known to have a higher ratio of SFR to gas surface densities 
$\Sigma_{\rm sfr}/\Sigma_{\rm gas}$ \citep{kennicutt98} -- have systematically higher ratios of high-density to total gas mass \citep{gao:2004.hcn.sfr.relation,
narayanan:2005.co32.lirgs,evans:agn.host.sfr,bussmann:2008.hcn32.sfr,
wu:2010.densemass.vs.lir}. 
The systematic increase of this ratio with surface density $\Sigma_{\rm gas}$ has been proposed as the origin of the difference between the apparently linear SFR-dense gas mass relation (which follows if the mass at a specific high-density has a high, constant star formation efficiency) and the super-linear 
Kennicutt-Schmidt relation (\citealt{gao:2004.hcn.sfr.relation,wu:2005.hcn.properties,krumholz:2007.dense.gas.tracer.schmidt}, 
but see also \citealt{yao:2003.molec.corr.vs.indicator,narayanan:2008.molec.sfr.indicators,narayanan:2008.sfr.densegas.corr}).

Exactly what determines the amount of dense gas, and hence the SFR, remains unknown. 
In simulations without stellar feedback, GMCs experience runaway collapse to densities much higher than observed, and rapidly turn a near-unity fraction of their gas into stars 
\citep{hopkins:rad.pressure.sf.fb,tasker:2011.photoion.heating.gmc.evol,
bournaud:2010.grav.turbulence.lmc,dobbs:2011.why.gmcs.unbound,
krumholz:2011.rhd.starcluster.sim,harper-clark:2011.gmc.sims}. 
Neither thermal pressure nor turbulence in and of itself can stave off collapse in the ISM: cooling is rapid, so that thermal support is quickly lost, and turbulent support  dissipates on a single crossing time (e.g., \citealt{ostriker:2001.gmc.column.dist}).  Some mechanism must therefore continuously inject energy and momentum into the gas on both GMC and galactic scales, in order to prevent runaway collapse to arbitrarily high densities. 

Various physical mechanisms have been proposed as a source of random motions in GMCs: photo-ionization, stellar winds, radiation pressure from UV and IR photons, proto-stellar jets, cosmic rays, supernovae, and gravitational cascades from large scales \citep[e.g.][and references therein]{mac-low:2004.turb.sf.review}. 
In \citet{hopkins:rad.pressure.sf.fb} (hereafter \paperone) and \citet{hopkins:fb.ism.prop} (\papertwo), 
we therefore developed a library of numerical hydrodynamic simulations of galaxies, with 
pc-scale resolution, molecular cooling, and explicit spacial/time resolution of 
feedback mechanisms including radiation pressure in the UV and IR, 
supernovae, massive and AGB stellar winds, and HII photo-ionization heating.
In \paperone\ and \papertwo, we show that these feedback mechanisms generically 
lead to the formation of a self-regulating, quasi steady-state multiphase ISM, in which 
dense GMCs form via gravitational collapse. Gas inside these GMCs then forms parsec scale clumps at densities $n>10^{4}\,{\rm cm^{-3}}$ in which most stars form; the GMCs hosting these clumps are then quickly broken up by feedback after they turn a few percent of their mass into stars. 

We showed in \papertwo\ that {most} properties of the ISM and GMCs 
are insensitive to the strength and precise mechanisms of feedback, so long as {\em sufficient} feedback is present to resist runaway dissipation and collapse. But this is largely a property of the low-density (non-star forming) gas in GMCs -- their properties are set to be those of any marginally self-gravitating object, i.e.\ they trace the Jeans mass and collapse conditions \citep{hopkins:excursion.ism}. The gas at high densities can, in principle, evolve very far away from its ``initial'' conditions even in just a couple GMC dynamical times. In this paper, we investigate the consequences of different feedback mechanisms for the dense gas in galaxies and GMCs. 


\vspace{-0.5cm}
\section{The Simulations}
\label{sec:sims}

The simulations used here are described in detail in 
\paperone\ (Sec.~2 \&\ Tables~1-3) and \papertwo\ (Sec.~2).
We briefly summarize the most important properties here. 
The simulations were performed with the parallel TreeSPH code {\small 
GADGET-3} \citep{springel:gadget}. They include stars, dark matter, and gas, 
with cooling, star formation, and stellar feedback. 

Gas follows an atomic cooling curve with additional fine-structure 
cooling to $<100\,$K, with no ``cooling floor'' imposed. 
Star formation is allowed only in dense regions above $n>1000\,{\rm cm^{-3}}$, 
at a rate $\dot{\rho}_{\ast}=\epsilon_{\ast}\,\rho_{\rm mol}/t_{\rm ff}$ 
where $t_{\rm ff}$ is the free-fall time, $\rho_{\rm mol}=f_{\rm H_{2}}\,\rho$ 
is the molecular gas density, and $\epsilon_{\ast}=1.5\%$ 
is a nominal efficiency at these densities 
\citep[][]{krumholz:sf.eff.in.clouds}. We follow \citet{krumholz:2011.molecular.prescription} 
to calculate the molecular fraction $f_{\rm H_{2}}$ in dense gas as a function 
of local column density and metallicity. 
In \paperone\ and \papertwo\ we show that the galaxy structure and SFR 
are basically independent of the small-scale SF law (independent of $\epsilon_{\ast}$ in particular), density threshold (provided it is high), 
and treatment of molecular chemistry. However, we discuss below how the properties of the most dense gas depend on these prescriptions. 

Stellar feedback is included, via a variety of mechanisms.

(1) {\bf Local Momentum Flux} from Radiation Pressure, 
Supernovae, \&\ Stellar Winds: Gas within a GMC (identified 
with an on-the-fly friends-of-friends algorithm) receives a direct 
momentum flux from the stars in that cluster/clump. 
The momentum flux is $\dot{P}=\dot{P}_{\rm SNe}+\dot{P}_{\rm w}+\dot{P}_{\rm rad}$, 
where the separate terms represent the direct momentum flux of 
SNe ejecta, stellar winds, and radiation pressure. 
The first two are directly tabulated for a single stellar population as a function of age 
and metallicity $Z$ and the flux is directed away from the stellar center. 
Because the local momentum flux is, by definition in our simulations,  interior to clouds, the systems are always optically thick to ultraviolet radiation, so  
the radiation force is approximately $\dot{P}_{\rm rad}\approx (1+\tau_{\rm IR})\,L_{\rm incident}/c$, 
where $1+\tau_{\rm IR} = 1+\Sigma_{\rm gas}\,\kappa_{\rm IR}$ accounts 
for the absorption of the initial UV/optical flux and multiple scatterings of the 
IR flux if the region is optically thick in the IR (with $\Sigma_{\rm gas}$ calculated 
for each particle given its location in the GMC). 

(2) {\bf Supernova Shock-Heating}: Gas shocked by 
supernovae can be heated to high temperatures. 
We tabulate the SNe Type-I and Type-II rates from 
\citet{mannucci:2006.snIa.rates} and STARBURST99, respectively, as a function of age and 
metallicity for all star particles and stochastically determine at 
each timestep if a SNe occurs. If so, the appropriate mechanical luminosity is 
injected as thermal energy in the gas within a smoothing length (nearest 32 gas neighbors) of the star particle. 

(3) {\bf Gas Recycling and Shock-Heating in Stellar Winds:} Gas mass is returned 
to the ISM from stellar evolution, at a rate tabulated from SNe and stellar mass 
loss (integrated fraction $\approx0.3$). The SNe heating is described above. Similarly, stellar winds 
are assumed to shock locally and inject the appropriate tabulated mechanical 
luminosity $L(t,\,Z)$ as a function of age and metallicity into the gas within a smoothing length. 

(4) {\bf Photo-Heating of HII Regions}: We also tabulate the rate of production of ionizing photons for 
each star particle; moving radially outwards from the star, we then ionize each neutral gas particle (using 
its density and state to determine the necessary photon number) 
until the photon budget is exhausted. Ionized gas is maintained at a minimum $\sim10^{4}\,$K until 
it falls outside an HII region.

(5) {\bf Long-Range Radiation Pressure:} Photons which escape the local GMC (not 
accounted for in mechanism (1) above) can be absorbed at larger radii. Knowing the intrinsic SED of each star 
particle, we attenuate integrating the local gas density and gradients to convergence. 
The resulting ``escaped'' SED gives a flux that propagates to large distances, and 
can be treated in the same manner as the gravity tree to give the local net incident flux 
on a gas particle. The local absorption is then calculated integrating over a frequency-dependent 
opacity that scales with metallicity, and the radiation pressure force is imparted. 

In implementing (1)-(5), all energy, mass, and momentum-injection rates are taken  from  stellar 
population models  \citep{starburst99}, assuming a \citet{kroupa:imf} IMF, without any free parameters.
More details, numerical tests, and resolution studies for these models are discussed in \papertwo; some additional numerical tests are discussed in Appendix~\ref{sec:appendix}.

We implement the model in four distinct initial disk models spanning a range of galaxy types. 
Each has a bulge, stellar and gaseous disk, halo, and central BH (although to isolate the 
role of stellar feedback, models for BH growth and feedback are disabled). 
At our standard resolution, each model has $\sim 0.3-1\times10^{8}$ total particles, 
giving particle masses of $500-1000\,\msun$ and $1-5$\,pc smoothing lengths, 
and are run for a few orbital times each. A couple ultra-high resolution runs for 
convergence tests employ $\sim10^{9}$ particles with sub-pc resolution.
The disk models include: 

(1) SMC: an SMC-like dwarf, with baryonic mass $M_{\rm bar}=8.9\times10^{8}\,\msun$ 
(gas $m_{g}=7.5\times10^{8}\,\msun$, bulge $M_{b}=1.3\times10^{8}\,\msun$, 
the remainder in a stellar disk $m_{d}$) and halo mass $M_{\rm halo}=2\times10^{10}\,\msun$. 
The gas (stellar) scale length is $h_{g}=2.1\,$kpc ($h_{d}=0.7$).

(2) MW: a MW-like galaxy, with 
halo $M_{\rm halo}=1.6\times10^{12}$, 
and baryonic 
$(M_{\rm bar},\,m_{b},\,m_{d},\,m_{g})=(7.1,\,1.5,\,4.7,\,0.9)\times10^{10}\,\msun$ 
with scale-lengths 
$(h_{d},\,h_{g})=(3.0,\,6.0)\,{\rm kpc}$. 

(3) Sbc: a gas-rich dwarf starburst disk with 
halo $M_{\rm halo}=1.5\times10^{11}$, 
and baryonic 
$(M_{\rm bar},\,m_{b},\,m_{d},\,m_{g})=(10.5,\,1.0,\,4.0,\,5.5)\times10^{9}\,\msun$ 
with scale-lengths 
$(h_{d},\,h_{g})=(1.3,\,2.6)\,{\rm kpc}$. 

(4) HiZ: a high-redshift massive starburst disk (typical of intermediate SMGs 
at $z\sim2-4$); $M_{\rm halo}=1.4\times10^{12}\,\msun$ (scaled for $z=2$ halos), 
and baryonic 
$(M_{\rm bar},\,m_{b},\,m_{d},\,m_{g})=(10.7,\,0.7,\,3,\,7)\times10^{10}\,\msun$ 
with scale-lengths 
$(h_{d},\,h_{g})=(1.6,\,3.2)\,{\rm kpc}$.

\begin{figure}
    \centering
    \plotonesize{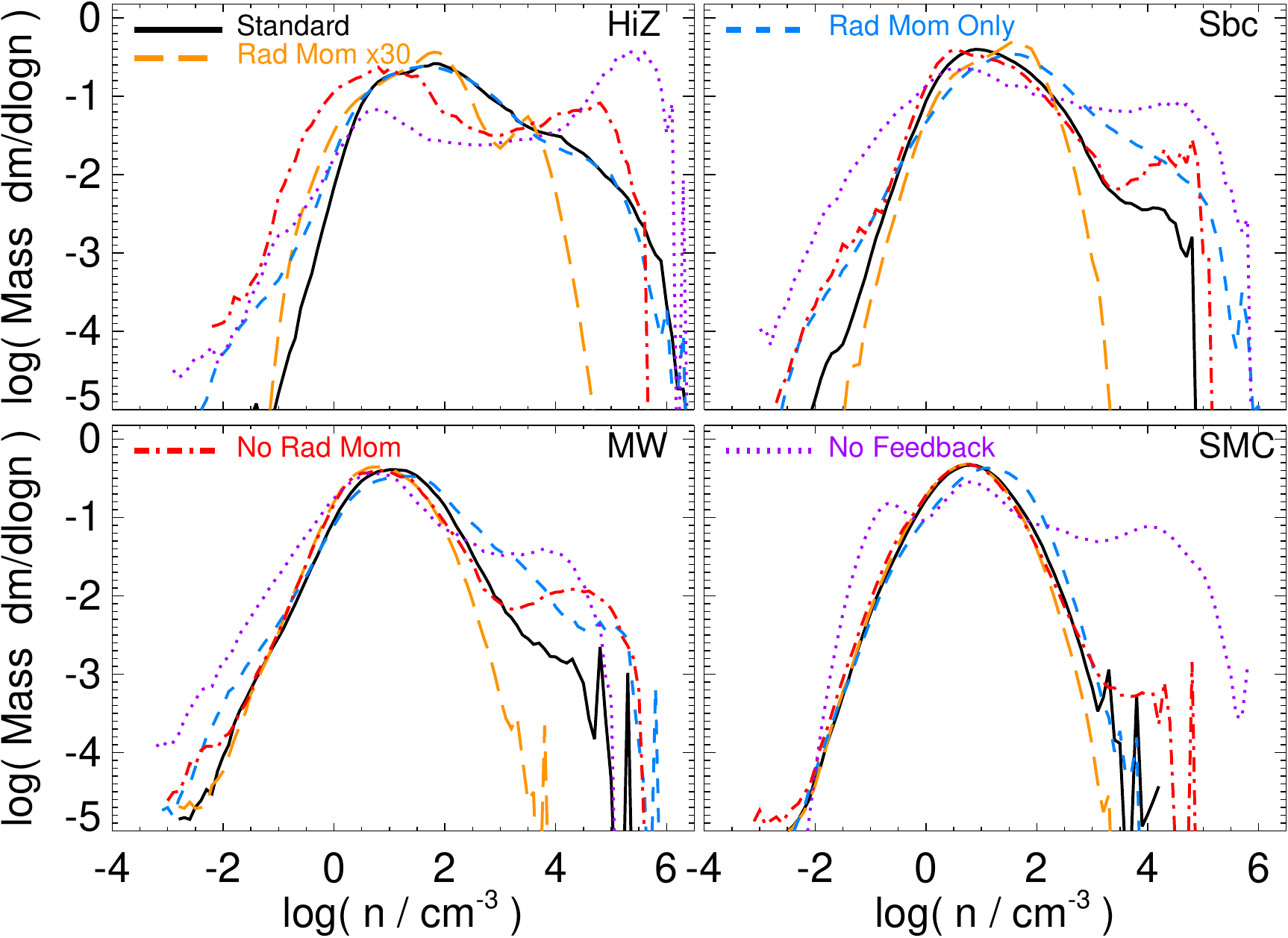}{0.99}
    \caption{Distribution of ``cold'' ($T<1000\,$K) gas densities $n$ in 
    each of the different disk galaxy models, with different feedback 
    mechanisms (\S~\ref{sec:sims}) enabled.
    ``Standard'' means all mechanisms are included. 
    In the ``no feedback'' case, gas piles up at $n\gg 10^{4}\,{\rm cm^{-3}}$.
    Even though SNe can regulate the {\em global} galaxy properties, 
    including just thermal feedback mechanisms (SNe \&\ stellar wind shock-heating 
    and HII photo-ionization) without radiation pressure or momentum flux 
    (``no momentum flux'') leads to a similar excess at high-$n$.
    Removing the heating mechanisms while keeping the radiation pressure 
    (``rad mom only'')
    has a modest effect on the high-$n$ gas, even though it can 
    dramatically change the galaxy wind and thermal state.
    If we make feedback stronger by simply increasing the strength of the radiation pressure momentum 
    flux by a factor of $30$, the amount of high-$n$ material is strongly suppressed.
    Remarkably, these changes have almost no effect on the median densities 
    of cold gas ($n\sim10-100\,{\rm cm^{-3}}$) -- i.e.\ typical densities of gas in GMCs -- 
    or the corresponding GMC mass function \&\ linewidths (nor do they much alter the 
    galaxy gas velocity dispersion or disk scale height; see \papertwo). It is the 
    dense gas that traces the effects of feedback.
    \label{fig:rho.dist.all}}
\end{figure}

\begin{figure*}
    \centering
    \plotsidesize{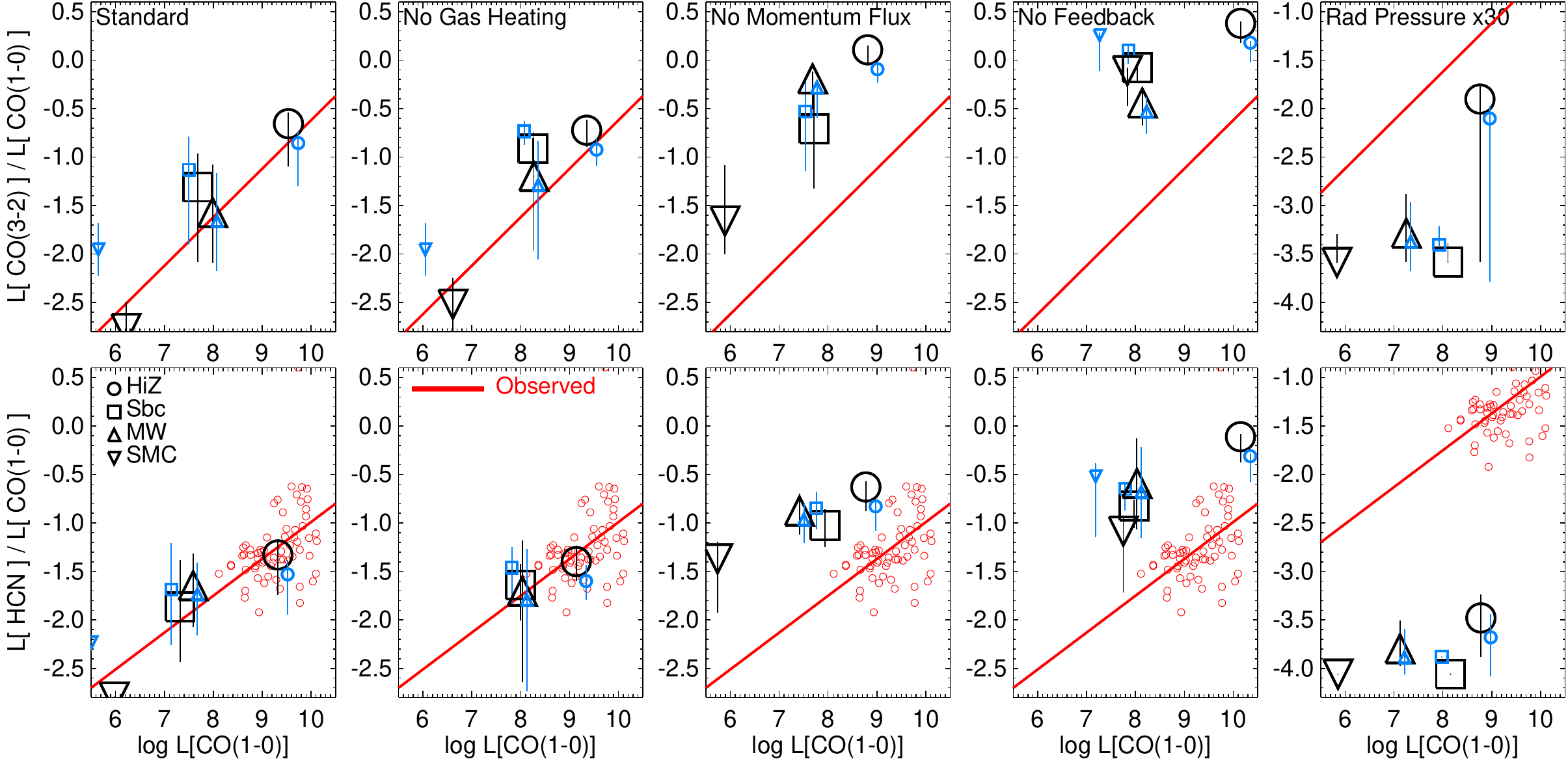}{0.95}
    \caption{Dependence of observational dense-gas tracers on 
    feedback and galaxy properties. 
    We show the predicted ratio of the luminosity in 
    HCN(1-0) ({\em bottom}) and CO(3-2) ({\em top}), 
    with critical densities $n_{\rm crit}\sim6\times10^{4},\,1.5\times10^{4}\,{\rm cm^{-3}}$, 
    to the luminosity in CO(1-0), which traces the total molecular/GMC 
    mass ($n\gtrsim100\,{\rm cm^{-3}}$). This is a proxy for the 
    ratio of very dense star-forming gas, to total GMC mass, as a function of 
    gas mass/SFR. The median and scatter in time for each galaxy model 
    is plotted (points with error bars). Large black points adopt a constant $\alpha_{\rm CO}$, 
    small blue points adopt $\alpha_{\rm CO}(\Sigma_{\rm H_{2}},\,Z)$ from \citet{narayanan:2011.xco.model}.
    The observed best-fit trend (line) and data points (small circles) are taken
    from \citet{gao:2004.hcn.compilation}. 
    We compare models with different feedback mechanisms, as 
    Fig.~\ref{fig:rho.dist.all}: 
    our standard model; no gas heating from SNe, stellar 
    winds, or photo-heating; no radiation pressure 
    momentum flux; no feedback; 
    and radiation pressure boosted by $30$.
    While global average GMC properties are similar in these models, 
    the ratio of very dense to GMC gas mass, and hence the relative 
    luminosity in tracers like HCN, are sensitive to feedback. 
    Gas heating has little effect on HCN because cooling times are so 
    short at these densities, but adjusting the strength of radiation pressure 
    leads to a nearly linear scaling in the predicted HCN luminosities. 
    In our standard models, the predictions agree well with observations, 
    both in the typical magnitude $L_{\rm HCN}/L_{\rm CO(1-0)}$ 
    (a consequence of ``normal'' feedback strength), 
    and the trend with $L_{\rm CO(1-0)}$, which stems from 
    rising surface densities in more luminous systems. 
    \label{fig:highrho.tracers}}
\end{figure*}

\begin{figure}
    \centering
    \plotonesize{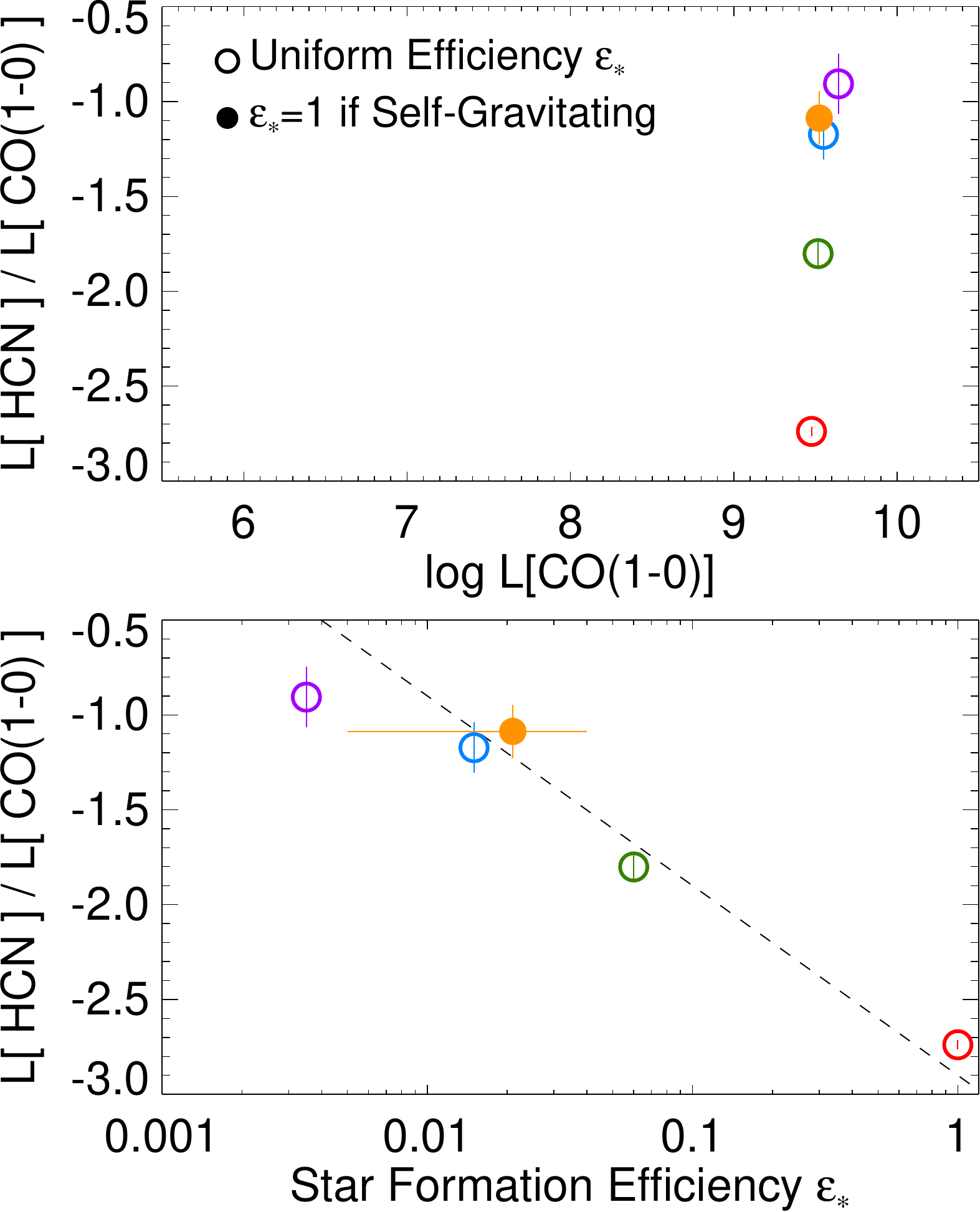}{0.95}
    \caption{Dependence of dense-gas tracers on the small-scale star formation efficiency $\epsilon_{\ast}$. We show results for a series of otherwise identical simulations of the HiZ model, as in Fig.~\ref{fig:highrho.tracers}, but systematically vary the assumed simulation star formation efficiency $\epsilon_{\ast}$ in the dense gas ($n>1000\,{\rm cm^{-3}}$), from which the SFR is $\dot{\rho}_{\ast} = \epsilon_{\ast}\,\rho_{\rm mol}/t_{\rm ff}$ (see \S~\ref{sec:sims}). {\em Top:} Very dense to GMC gas mass ratio as a function of GMC gas mass (as Fig.~\ref{fig:highrho.tracers}). {\em Bottom:} Same HCN(1-0) to CO(1-0) ratio, for the same simulations, as a function of the star formation efficiency $\epsilon_{\ast}$ assumed. The solid point does not assume a constant $\epsilon_{\ast}$, but adopts the model in \citet{hopkins:binding.sf.prescription}: the efficiency is $\epsilon_{\ast}=1$ in regions which are locally self-gravitating, but $\epsilon_{\ast}=0$ otherwise. We therefore plot the time and mass-averaged efficiency $\langle \epsilon_{\ast} \rangle$ predicted (with its scatter), around $\sim0.5-4\%$. 
    As shown in \paperone, the total SFR, total IR luminosity, and here, total CO luminosity (GMC gas mass) are nearly identical (insensitive to the small-scale SF law), because it is set by the SFR needed to balance collapse via feedback. However, to achieve the same SFR with lower (higher) efficiency, a larger (smaller) mass of dense gas is needed. The dense-to-GMC gas ratio scales approximately inversely with the mean $\langle \epsilon_{\ast} \rangle$ (dashed line shows a fit with slope $L[{\rm HCN}]/L[{\rm CO(1-0)}] \propto \langle \epsilon_{\ast} \rangle^{-1}$). 
    \label{fig:highrho.tracers.sfe}}
\end{figure}

\vspace{-0.5cm}
\section{High-Density Gas \&\ Feedback}
\label{sec:extreme.rho}

In Figure~\ref{fig:rho.dist.all}, we plot the distribution of cold gas 
($T<1000$\,K, mostly in GMCs) densities in each simulation. 
We show this for models with various feedback mechanisms enabled 
or disabled in turn. The basic properties of the 
``standard'' model (all feedback enabled) are discussed in 
\papertwo: the distribution has a lognormal core (dispersion $\sim1-1.5$\,dex) 
with median $\langle n \rangle\sim 100\,{\rm cm^{-3}}$, 
in good agreement with observations. 
As discussed in \S~\ref{sec:intro}, the behavior of this ``core'' is not 
very sensitive to the model -- they are just set by the conditions for 
gravitational collapse in a turbulent ($Q\sim1$) disk \citep[see][]{hopkins:excursion.ism}. 
The {\em average} GMC properties (sizes, linewidths, densities, etc) are 
not sensitive tracers of feedback. 

But the high-$n$ tail shows significant differences between models.
Turning off feedback entirely leads to runaway collapse 
with a large secondary peak at $n\rightarrow10^{6}\,{\rm cm^{-3}}$ 
(the maximum resolved density). 
Turning off ``heating'' mechanisms (shock-heating by SNe and stellar winds 
and photo-ionization heating) leads to a small increase in the 
amount of high-density $n\gtrsim10^{4}\,{\rm cm^{-3}}$ 
material in the MW and SMC models. 
But in all models, even the MW and SMC cases -- where 
we show in \papertwo\ that SNe heating may dominate the {\em global} self-regulation 
of the disk and generation of galactic winds \citep{hopkins:stellar.fb.winds} -- 
turning off radiation pressure (while still keeping SNe, stellar wind, and photo-heating) 
yields a much more dramatic increase in the amount of very dense material. 
In the HiZ and Sbc models, this is nearly as 
significant as turning off all feedback. 
In other words, even where global self-regulation can be set by SNe heating, the dense material 
at $n\gtrsim 10^{4}\,{\rm cm^{-3}}$ is regulated by radiation 
pressure. This should be expected: at these densities optical depths are 
large in the infrared, and cooling times for SNe remnants are $\sim10^{4}$ times shorter than 
the dynamical time.

We illustrate how this very dense gas can function as a tracer of the 
strength of radiation pressure by also comparing a series of 
otherwise identical models where we simply multiply the local radiation pressure force 
applied by a large, constant factor $\eta_{p}=30$ everywhere in the simulation. 
Again, the median $\langle n \rangle$ is similar in both cases. 
However, to maintain {global} equilibrium, the boosted-feedback model requires 
much lower star formation rates (and mass in young stars). 
In turn, the high-$n$ tail (where stars actually form) is much smaller.

\vspace{-0.5cm}
\section{Comparison With Observations}
\label{sec:rho.high.tracers}

In Figure~\ref{fig:highrho.tracers}, we consider what these differences 
mean in terms of observational tracers. The total mass budget in intermediate-density 
molecular gas, roughly gas with $n\gtrsim100\,{\rm cm^{-3}}$, is commonly traced by the CO(1-0) transition. 
Since we know the mass in gas above this density, we can use the observed relation 
$M_{\rm mol}\approx 4.8\,L_{\rm CO(1-0)}/({\rm K\,km\,s^{-1}\,pc^{2}})$ 
\citep{youngscoville:1991.mol.gas.review} 
to estimate $L_{\rm CO}$. Similarly, the HCN luminosity $L_{\rm HCN}$ is related 
to the mass above its critical density, $n_{\rm crit}\approx6\times10^{4}\,{\rm cm^{-3}}$, 
as $M(n>n_{\rm crit})\approx 5.5\,L_{\rm HCN}/({\rm K\,km\,s^{-1}\,pc^{2}})$ 
\citep{gao:2004.hcn.compilation}.
Another high-density tracer is CO(3-2), 
with $n_{\rm crit}\approx 1.5\times10^{4}\,{\rm cm^{-3}}$ and 
$M(n>n_{\rm crit})\approx 1.33\,L_{\rm CO(3-2)}/({\rm K\,km\,s^{-1}\,pc^{2}})$ 
\citep{narayanan:2005.co32.lirgs}. 
All of these conversions are uncertain at the factor of $\sim2$ 
level, but a more accurate prediction would require a 
full non-LTE radiative transfer solution (outside the 
scope of this paper), and in any case, such a systematic 
offset does not change our conclusions. For example, we contrast the results with the choice of the surface-density and metallicity-dependent $\alpha_{\rm CO}$ derived in \citet{narayanan:2011.xco.model}, and obtain similar results.\footnote{It is also not simply the case that these lines just trace the mass above a critical density. \citet{evans:1999.sf.gmc.review} advocate, for example, using instead the ``effective'' density of tracers (the density needed to produce a line with radiation temperature of $1\,$K). We can re-consider our calculations using these densities, but we must be careful to also use the appropriate mass-to-light conversion calibrated for virial masses at these densities instead. From \citet{wu:2010.densemass.vs.lir}, the effective density for HCN(1-0) is $n_{\rm eff} \approx 5.1\times10^{3}\,{\rm cm^{-3}}$, which for our standard SMC, MW, Sbc, and HiZ models increases the total gas mass with $n>n_{\rm eff}$ (relative to that with $n>n_{\rm crit}$) by a factor of $(1.7,\,1.9,\,4.5,\,4.1)$, respectively. However the authors also re-calibrate the appropriate mass-to-light ratio (see their \S~5.1), giving a best fit $L_{\rm HCN}/M(n>n_{\rm eff})$ conversion factor which is a factor $\approx 0.3$ times our adopted $L_{\rm HCN}/M(n>n_{\rm crit})$. As a result, the predicted $L_{\rm HCN}$ is systematically different by only $\approx50\%$ (slightly larger for the Sbc and HiZ models, smaller for the SMC and MW models). We see similar small changes in the simulations with varied feedback physics.}

Because we cannot follow gas to arbitrarily high densities where star formation 
actually occurs, the simulations enforce the observed relation between 
dense gas mass and star formation efficiency (and indeed most of the star 
formation occurs in gas with resolved densities $\gtrsim10^{4}\,{\rm cm^{-3}}$). 
This means they automatically reproduce the observed 
$L_{\rm HCN}-$SFR or $L_{\rm HCN}-L_{\rm IR}$ relations: they define our 
star formation prescription.\footnote{If we arbitrarily raise/lower the efficiency of star formation in the dense gas, we will systematically shift the amount of dense gas required for the SFR and corresponding feedback strength needed to self-regulate while preserving otherwise similar galaxy properties (this is shown explicitly in Figure~8 of \paperone). Hence this would also alter the density PDFs and $L_{\rm HCN}-L_{\rm CO}$ relations in Figures~\ref{fig:rho.dist.all}-\ref{fig:highrho.tracers}, but at the expense of manifestly violating the observed $L_{\rm HCN}-$SFR and $L_{\rm HCN}-L_{\rm IR}$ relations. We investigate this in more detail below.}

However, Figure~\ref{fig:rho.dist.all} shows 
that the relative amount of high-density ($n\gtrsim10^{4}\,{\rm cm^{-3}}$) 
and intermediate-density/total molecular ($n\gtrsim 100\,{\rm cm^{-3}}$) gas 
can vary dramatically.
The relevant diagnostic is therefore the mass fraction in 
GMCs that is in dense cores, $M_{\rm dense}/M_{\rm tot}$, 
traced in $L_{\rm HCN}/L_{\rm CO(1-0)}$ 
or $L_{\rm CO(3-2)}/L_{\rm CO(1-0)}$. 
Figure~\ref{fig:highrho.tracers} plots this as a function of total cold gas mass $L_{\rm CO(1-0)}$. 
For both tracers, we compare the observed relations:
\begin{align}
\frac{L_{\rm HCN}}{L_{\rm CO(1-0)}} & \propto 
0.1\,{\Bigl (}\frac{L_{\rm CO(1-0)}}{10^{10}\,{\rm K\,km\,s^{-1}\,pc^{-2}}}{\Bigr)}^{0.4} \\ 
\frac{L_{\rm CO(3-2)}}{L_{\rm CO(1-0)}} & \propto 
0.1\,{\Bigl (}\frac{L_{\rm CO(1-0)}}{10^{10}\,{\rm K\,km\,s^{-1}\,pc^{-2}}}{\Bigr)}^{0.5} 
\end{align}
\citep{gao:2004.hcn.compilation,narayanan:2005.co32.lirgs}.

In the ``standard'' (all feedback enabled) models, the 
predicted ratios of dense-to-cold gas and the scaling with gas mass/luminosity/SFR 
agree remarkably well with observations.

\vspace{-0.5cm}
\section{What Determines Dense Core Fractions \&\ HCN Luminosities?}
\label{sec:rho.high.tracers:overview}


How does this scaling arise in our simulations? 
Dense clumps form and collapse until they produce sufficient stars 
such that the local feedback can offset gravity and disrupt the clump (generate a force 
$F_{\rm feedback}\sim G\,M_{\rm cl}^{2}/R^{2}\sim G\,M_{\rm cl}\,\Sigma_{\rm cl}$).
The total force needed to unbind the GMC is then $F_{\rm tot}\sim G\, M_{\rm GMC}\,\Sigma_{\rm GMC}$. 
If all the feedback acting on the GMC is 
from young stars currently in dense clumps (and coupled therein), then 
the total force is simply the sum over the forces acting in each 
dense region $i$, $F_{\rm tot}=\sum{F_{\rm dense,\,i}}
=G\,\sum{M_{\rm dense,\,i}\,\Sigma_{\rm dense,\,i}} 
= G\,M_{\rm dense,\,tot}\,\langle \Sigma_{\rm dense} \rangle$. 
Equating the two gives 
$M_{\rm dense}/M_{\rm GMC}\sim \Sigma_{\rm GMC}/\langle \Sigma_{\rm dense} \rangle$.

More generally, the stars form in dense regions, but these can have a lifetime 
which is short compared to the massive stellar evolution timescale ($\sim 5\,$Myr). 
For clumps that live for $N_{t}\sim1$ free-fall times, 
the fraction of the GMC luminosity in dense clumps is 
$f_{L}\approx 0.1\,n_{4}^{-1/2}\,N_{t}$, where 
$n_{4}\equiv \langle n_{\rm cl} \rangle/10^{4}\,{\rm cm^{-3}}$.\footnote{If the massive stars dominating the luminosity have lifetime $t_{\ast}\approx 5\,$Myr, and clumps live for a time $t_{\rm cl} \equiv N_{t}\,t_{\rm ff} = N_{t}/\sqrt{(32/3\pi)\,G\,\rho_{\rm cl}} \approx 0.5\,N_{t}\,(n_{\rm cl}/10^{4}\,{\rm cm^{-3}})^{-1/2}$\,Myr, then (on average) if $t_{\rm cl}<t_{\ast}$ the fraction of massive stars (light) in clumps is $f_{L}\sim t_{\rm cl}/t_{\ast}\sim0.1\,n_{4}^{-1/2}\,N_{t}$.} 
As Figure~\ref{fig:rho.dist.all} shows, the dense gas is most affected 
by the local radiation pressure. The total force (momentum 
deposition rate) from radiation pressure 
in a (smooth) clump with average optical depth $\tau$ in the IR 
is $\dot{p}=(1+\tau)\,L/c$; the total $\dot{p}$ in the 
GMC is then $(1+[\tau_{\rm dense}-\tau_{0}]\,f_{L}+\tau_{0})\,L/c$ 
(where $L$ is the total stellar luminosity, $\tau_{0}=\kappa\,\Sigma_{\rm GMC}$ is the 
mean $\tau$ of the GMC as a whole, 
and $\tau_{\rm dense}=\kappa\,\Sigma_{\rm dense}$ is the mean $\tau$ of a dense 
clump).\footnote{There is some debate in the literature regarding the exact form of the $\eta\sim(1+\tau)$ prefix in the momentum flux $\eta\,L/c$ (see e.g.\ \citealt{krumholz:2012.rad.pressure.rt.instab}, but also \citealt{kuiper:2012.rad.pressure.outflow.vs.rt.method}). For our purposes in this derivation, it is simply an ``umbrella'' term which should include all momentum flux terms (including radiation pressure in the UV and IR, stellar winds, cosmic rays, warm gas pressure from photo-ionization/photo-electric heating, and early-time SNe). These other mechanisms will introduce slightly different functional dependencies in our simple derivation, if they are dominant, however, the total value of $\eta\sim$a few is likely to be robust even if IR radiation pressure is negligible, leading to the same order of magnitude prediction. And we find that using a different form of the radiation pressure scaling which agrees quite well with that calculated in \citet{krumholz:2012.rad.pressure.rt.instab} has only weak effects on our results (see \paperone, Appendix~B, \papertwo, Appendix~A2, \&\ \citealt{hopkins:clumpy.disk.evol}, Appendix~A).} 
We can estimate the required $L_{\rm cl}/M_{\rm cl}$ 
for each dense clump by again equating the force to gravity, 
and then $L=f_{L}^{-1}\,\sum{L_{\rm cl}}$. Using this and equating 
the total force on the GMC to its self-gravity (and assuming 
$\tau_{\rm dense}\gtrsim1\gg \tau_{0}$), 
we obtain
\begin{align}
\label{eqn:mdense}
\frac{M_{\rm dense}}{M_{\rm GMC}} &\approx 
\frac{f_{L}\,\tau_{0}}{1+f_{L}\,\tau_{\rm dense}+\tau_{0}} \\ 
&\approx
\frac{0.03\,N_{t,3}\,n_{4}^{-1/2}\,\Sigma_{\rm GMC,\,100}}
{1 + 0.3\,N_{t,3}\,n_{4}^{-1/2}\,\Sigma_{\rm dense,\,1000} + 0.1\,\Sigma_{\rm GMC,\,100}} \\ 
&\rightarrow
\frac{\Sigma_{\rm GMC}}{\Sigma_{\rm dense}}\ \ \ \ \ (n_{4}^{-1/2}\,\Sigma_{\rm dense,\,1000}\gg1)
\end{align}
where 
$N_{t,\,3}=N_{t}/3$, 
$\Sigma_{\rm GMC,\,100}\equiv \Sigma_{\rm GMC}/100\,\msun\,{\rm pc^{-2}}$ 
($\sim1$ for typical GMCs, i.e., in non-ULIRGs), 
$\Sigma_{\rm dense,\,1000}\equiv \Sigma_{\rm dense}/1000\,\msun\,{\rm pc^{-2}}$ 
($\sim1$ for typical dense clumps),
and $n_{4}\equiv n_{\rm dense}/10^{4}\,{\rm cm^{-3}}$. 

In simulations, we find typical $N_{t}\approx2-4$; if we assume a fixed 
SFR per free-fall time $\dot{M}_{\ast}=\epsilon_{\ast}\,M_{\rm dense}/t_{\rm free-fall,\,dense}$ 
and use the above derivation to obtain the critical $L/M$ in a dense clump, 
we predict $N_{t}\approx3\,(0.05/\epsilon_{\ast})$. 
These values agree well with observational 
estimates \citep[][and references therein]{evans:2009.sf.efficiencies.lifetimes}. 
Thus up to some ``saturation'' level when $n_{4}^{-1/2}\,\Sigma_{\rm dense,\,1000}\gg1$, we expect $M_{\rm dense}/M_{\rm GMC} \propto \epsilon_{\ast}^{-1}$ (inversely proportional to the small-scale star formation efficiency, if it is constant); we demonstrate this explicitly below. 

Note that one might also expect a dependence on metallicity, 
since the opacities $\tau$ appear; however, accounting for it properly (assuming 
opacity scales linearly with metallicity), the dependence on metallicity cancels 
nearly completely.

The predicted ratio $M_{\rm dense}/M_{\rm GMC}$ 
increases with $\Sigma_{\rm GMC}$. We saw in \papertwo\ that 
$\Sigma_{\rm GMC}$ increases with average galaxy surface density (hence galaxy SFR and 
luminosity). This drives the trend of increasing $L_{\rm HCN}/L_{\rm CO(1-0)}$ 
at higher luminosities in Figure~\ref{fig:highrho.tracers}. 
Observationally, from the baryonic Tully-Fisher relation ($R_{g}\propto M_{g}^{0.25-0.33}$) 
\citep[e.g.][]{stark:baryonic.tg.in.gas.rich.gal}, we expect (for Jeans-scale clouds) average 
surface densities $\Sigma_{\rm GMC}\propto M_{g}^{0.3-0.5}$ (which fits well the 
direct estimates in \papertwo). 
This leads to the prediction that $L_{\rm HCN}/L_{\rm CO(1-0)} 
\propto L_{\rm CO(1-0)}^{(0.3-0.5)}$ for ``normal'' galaxies 
(with an upper limit when $M_{\rm dense}/M_{\rm GMC}\sim1$). 
Similar considerations can be used to derive the observed IR-CO(1-0) scaling \citep{andrews:obsd.gal.vs.edd.limit}. 

Note that this argument assumes that $\Sigma_{\rm dense}$ {\em does not} increase with average galaxy surface density, or at least not as rapidly as $\Sigma_{\rm GMC}$ does. There is some observational evidence that, at the highest clump masses, $\Sigma_{\rm dense}$ does not increase with increasing clump mass (e.g. Figure~8 in \citealt{murray:sizes.lum.star.clusters}).

This simple force argument also predicts, for example, that in 
local LIRGs and ULIRGs, where enhanced star formation is driven 
by extremely dense nuclear concentrations of gas, and so 
$\Sigma_{\rm GMC}$ {\em must} be large (it must be 
at least the mean density), the dense gas fraction or 
$L_{\rm HCN}/L_{\rm CO(1-0)}$ will be systematically 
larger. This has been observed \citep{gao:2004.hcn.sfr.relation,
narayanan:2005.co32.lirgs,evans:agn.host.sfr,bussmann:2008.hcn32.sfr}. 
Moreover, we can estimate the magnitude of this enhancement: 
in typical ULIRG nuclei, where the effect will be most extreme, 
the surface densities reach $\Sigma\gtrsim10^{3}\,\msun\,{\rm pc^{-2}}$, 
reaching the limit where 
$L_{\rm HCN}/L_{\rm CO(1-0)}\sim M_{\rm dense}/M_{\rm GMC} 
\sim \langle \Sigma \rangle/\Sigma_{\rm dense} \sim 0.1-1$ 
(i.e.\ where the dense gas fractions saturate)
a factor $\sim10$ larger than that in normal galaxies. 
This agrees well with the enhancements in $L_{\rm HCN}$ 
observed for ULIRGs and dense  relative to normal spiral galaxies in \citet{gao:2004.hcn.sfr.relation,juneau:2009.enhanced.dense.gas.ulirgs}. 

Finally, note that our choice to normalize the above scalings around $\sim10^{4}\,{\rm cm^{-3}}$ is purely for convenience (since this is near the densities of interest for the tracers we discuss here). The simple scaling argument above admits a continuum of densities, with Eq.~\ref{eqn:mdense} applicable to a wide range of densities $n\gg n_{\rm GMC}$. If the SFR per dynamical time in the dense gas ($\epsilon_{\ast}$) is constant, then this simply predicts $M_{\rm dense}(n>n_{\rm crit})/M_{\rm GMC} \propto n_{\rm crit}^{-1/2}$ (up to the ``saturation level'' noted above).

\vspace{-0.5cm}
\section{Dense Tracers Versus Feedback}
\label{sec:rho.high.tracers:fb}

Based on the derivations above, we can guess how 
the HCN and CO luminosities will behave under various 
changes to the feedback model, shown in 
Figure~\ref{fig:highrho.tracers}. Because it has 
little effect on the most dense gas, removing gas 
heating has a weak effect on the 
predicted correlations, even in the MW and SMC 
models. Note that it can move systems {\em along} the correlations -- it 
does so by globally regulating the mass in winds, and so the total 
mass forming GMCs and ultimately forming stars (e.g.\ absolute 
$L_{\rm CO(1-0)}$). But within GMCs, 
heating has little effect, since the cooling time is so short.
Removing the local direct radiation pressure momentum flux,  
however, has a much more dramatic effect, similar to removing 
all forms of feedback entirely. In both cases, there is little or 
nothing to resist runaway collapse inside the GMCs, and dense gas 
piles up until it saturates at fractions of order unity relative 
to the total GMC mass (Figure~\ref{fig:rho.dist.all}).
These produce order-of-magnitude or larger discrepancies 
with the observations (in the SMC and MW case, momentum 
flux from SNe resist complete collapse, but the amount of dense 
gas is still $\sim10$ times too large for the observations). 
On the other hand, if we artificially 
boost the local radiation pressure strength by a 
large factor, the dense gas is all removed. 
If we repeat the derivations above, with the radiation pressure force 
multiplied uniformly by a factor $\eta$, 
the predicted $M_{\rm dense}/M_{\rm GMC}$ in the non-saturated 
regime ($\Sigma_{\rm GMC}\ll \Sigma_{\rm dense}$) simply 
scales $\propto \eta^{-1}$. This is what we find -- 
the predicted $L_{\rm HCN}/L_{\rm CO(1-0)}$ 
decreases approximately linearly in proportion to the 
``boost'' in feedback strength.

\vspace{-0.5cm}
\section{Dense Tracers Versus Star Formation Efficiencies}
\label{sec:rho.high.tracers:sfe}

From the scaling in \S~\ref{sec:rho.high.tracers:overview}, we can also anticipate how the HCN and CO luminosities will behave as we change the small-scale star formation law adopted in the simulations. This is shown explicitly in Fig.~\ref{fig:highrho.tracers.sfe}. Recall, as discussed above and shown in \paperone\ and \papertwo, changing the SF law in high-density gas has essentially no effect on the total SFR of the simulations. This is because star formation is feedback-regulated, so simply requires a balance between a certain number of young stars (hence total feedback input) and global collapse/dissipation. 

In Fig.~\ref{fig:highrho.tracers.sfe}, we systematically vary the star formation efficiency $\epsilon_{\ast}$ (SFR per dynamical time in dense gas with $n>1000\,{\rm cm^{-3}}$), by a factor of $\sim10^{3}$. We also consider a different model in which the instantaneous local SF efficiency is set to either unity or zero depending on whether a given gas parcel is locally self-gravitating (at the resolution limit), which produces a time and volume-averaged efficiency of $\langle \epsilon_{\ast}\rangle \sim 2\%$ but with large variability. Across all these models, we find a $\lesssim20\%$ change in the total time-averaged SFR. Here, we see similarly that there is also almost no effect on the intermediate-density gas, traced in $L_{\rm CO(1-0)}$. However, there is a strong systematic trend in the amount of high-density gas, reflected in $L_{\rm HCN}$. In order to globally self-regulate against runaway collapse, a certain total amount of feedback, hence total SFR, is needed; but to achieve the same SFR with systematically lower (higher) efficiency, a correspondingly larger (smaller) amount of dense gas must be present (see also \paperone, Fig.~5). The scaling is roughly inverse with efficiency, $L_{\rm HCN} \propto \epsilon_{\ast}^{-1}$. 

Thus, the ratio of high-density to intermediate-density gas, in tracers such as $L_{\rm HCN}/L_{\rm CO(1-0)}$, is really a direct tracer of the amount of feedback per ``unit dense gas,'' i.e.\ the product of the feedback and star formation efficiencies: $L_{\rm HCN}/L_{\rm CO(1-0)} \propto (\eta\,\epsilon_{\ast})^{-1}$.

However, at fixed feedback strength but changing $\epsilon_{\ast}$, recall that the total SFR (for otherwise identical galaxies) is the same, while the amount of dense gas changes. Hence a tracer of the ratio of dense gas to total SFR, e.g. $L_{\rm HCN}/L_{\rm IR}$, is able to independently constrain $\epsilon_{\ast}$ and break this degeneracy. Here, we find very good agreement between the predicted $L_{\rm HCN}/L_{\rm IR}$ in our ``default'' models with fixed $\epsilon_{\ast} = 0.015$, and/or in the model with variable (self-gravity dependent) efficiencies (since this produces a very similar average efficiency). Assuming fixed $\epsilon_{\ast}=1$, on the other hand, leads us to predict a mean ratio $L_{\rm HCN}/L_{\rm IR}$ a factor $\sim50$ lower than observed \citep[in e.g.][]{gao:2004.hcn.compilation,wu:2010.densemass.vs.lir}. 

\vspace{-0.5cm}
\section{Discussion}
\label{sec:discussion}

We have used a library of numerical hydrodynamic simulations of galaxies, with 
pc-scale resolution, molecular cooling, and explicit spacial/time resolution of 
feedback mechanisms including radiation pressure in the UV and IR, 
supernovae, massive and AGB stellar winds, and HII photo-ionization heating to 
study how the properties of dense gas -- i.e.\ the gas where stars actually form -- 
can be a sensitive tracer of the effects of feedback and a strong constraint on 
models of star formation and stellar feedback. 
In \paperone\ and \papertwo, we show that these feedback mechanisms generically 
lead to the formation of a self-regulating, quasi steady-state multiphase ISM, in which 
dense GMCs form via gravitational collapse. 
The GMCs then form clumps at densities 
$n>10^{4}\,{\rm cm^{-3}}$ in which most stars form. 
These stars, which have a total mass amounting to only a few percent of the GMC mass, 
then disrupt their parent GMCs. 
We showed in \papertwo, however, that {\em most} properties of the ISM and GMCs 
are insensitive to the feedback mechanism or strength of feedback, so long as 
some mechanisms are present that are sufficient to resist runaway dissipation and collapse 
in these GMCs. Considering models with e.g.\ the dominant feedback mechanism 
being radiation pressure, SNe, HII photo-heating, or stellar winds; or models with different 
density thresholds, power-law dependencies, or efficiencies of star formation in dense sub-clumps; 
or models where we arbitrarily multiply/divide the strength of feedback by large factors; we find 
that provided something can make GMCs short-lived, their global properties (mass functions, 
densities, size-mass and linewidth-size relations, virial parameters, and ISM velocity 
dispersions, scale-heights, phase structure, and Toomre Q parameters)
largely reflect global gravitational conditions rather than e.g.\ some local 
hydrostatic equilibrium that would be sensitive to the details of star formation and/or feedback 
on small scales. 

However, we show here that the properties of the very dense gas, $n\gtrsim10^{4}\,{\rm cm^{-3}}$,  
{\em are} sensitive to the strength and nature of feedback, and the star formation efficiency on small scales. 
If feedback is inefficient, then dense regions within GMCs will collapse 
and accrete until a large fraction of the GMC mass is in
dense clumps. If, on the other hand, feedback is efficient, then only a small fraction of the 
dense clumps within a cloud collapse before sufficient massive stars are formed to unbind 
the parent cloud. If star formation is inefficient, a ``bottleneck'' appears and more gas must pile up at these densities to ultimately produce the same strength of feedback.
The ratio of mass in dense gas -- traced by dense molecular transitions such as $L_{\rm HCN}$ -- 
to the total mass of cool gas -- traced by lower-density transitions such as $L_{\rm CO(1-0)}$ -- 
is therefore a sensitive measure of feedback and star formation efficiencies.
In models with weak or no feedback, we show that the predicted ratio $L_{\rm HCN}/L_{\rm CO(1-0)}$ 
is an order-of-magnitude or more larger than observed. 
In models with feedback efficiencies taken ``as is'' from stellar evolution models, 
the predicted ratio agrees well with that observed. 
But if we arbitrarily make feedback more or less strong, multiplying the momentum flux by some factor 
$\eta$, then the predicted $L_{\rm HCN}/L_{\rm CO(1-0)}$ ratio scales approximately as 
$\eta^{-1}$; a model with $30$ times stronger radiation pressure predicts $\sim30$ times lower 
$L_{\rm HCN}/L_{\rm CO(1-0)}$. 

Likewise, if the star formation efficiencies either follow a physically motivated model where they reflect rapid collapse in only self-gravitating regions, or an imposed average $\sim1\%$ per free-fall time in the dense gas, then the predicted ratios agree well with the observations, but if we make the star formation efficiency much lower (higher), the ratio of dense gas tracers to the total SFR ($L_{\rm HCN}/L_{\rm IR}$, for example) becomes much higher (lower) than observed. In some sense, then, the question of why star formation efficiencies are low in a galaxy-wide sense is shifted to the question of why most very high-density gas is not forming stars rapidly (which these simulations cannot predict). Whether this is a consequence of ``slow'' star formation therein, or most such gas not being self-gravitating, or some additional physics not included here, remains an important subject for future work.

The predicted ratios are sensitive not just to the absolute strength of feedback, but also to the 
mechanisms of feedback. The thermal energy deposited by SNe explosions, for example, 
can have dramatic effects on galaxy scales (puffing up galaxy disks, driving turbulence, and 
accelerating material into galactic super-winds), once it escapes from dense gas to efficiently heat 
low-density material. But it has almost no effect directly {\em within} the high-density gas, 
because the post-shock cooling time at these densities is $\sim10^{4}$ times shorter than the dynamical 
time. We therefore predict that this mechanism has little effect on the observed $L_{\rm HCN}$-$L_{\rm CO(1-0)}$ 
relation, except to move galaxies {\em along} the relation by globally regulating the amount of 
(previously low-density) gas which can cool into new GMCs.
Radiation pressure, on the other hand, can provide a very 
strong source of feedback at these densities, because these regions can be optically thick. 
Direct observations of massive star-forming regions have begun to suggest such a 
scenario \citep{lopez:2010.stellar.fb.30.dor}; correlating this with observations of the dense-to-low density 
gas mass ratio would provide a mechanism-specific diagnostic of feedback strength. 

In addition to the normalization (median ratio $L_{\rm HCN}/L_{\rm CO}$), 
we show also that the systematic trends in $L_{\rm HCN}-L_{\rm CO}$ with 
e.g.\ SFR, $L_{\rm IR}$, or $L_{\rm CO}$ are tracers of feedback properties. 
We predict that the ratio of dense to total gas increases with increasing 
SFR and luminosity, in a manner in good agreement with observations. 
This arises because higher-SFR systems tend to have higher cloud surface densities 
(equivalently, higher pressures) and therefore require 
more force (more 
star formation, for the same feedback efficiencies) to unbind. 
We therefore predict a correlation between the dense gas ratio tracers and 
cloud surface densities of the form $\propto \Sigma_{\rm GMC}$ (at low 
surface densities; it must saturate at higher densities), which should explain some of the observed scatter 
in nominal cloud lifetimes, dense gas formation efficiencies, and measured star 
formation efficiencies. This specifically predicts an enhanced ratio of $L_{\rm HCN}/L_{\rm CO(1-0)}$ 
in extreme systems such as ULIRGs and galaxy nuclei, and 
very high-redshift starburst disks -- these systems have disk-average surface densities 
$\Sigma$ in excess of the ``typical'' MW $\Sigma_{\rm GMC}$, so must have 
significantly higher $\Sigma_{\rm GMC}$. This should also manifest in even higher-density tracers such as 
HCN(3-2) \citep[see e.g.][]{bussmann:2008.hcn32.sfr}, for which our models here predict a more dramatic difference in the most extreme systems (but begins to push against our resolution limits). Such an 
enhancement has been tentatively observed; we 
provide a motivation for further observations to constrain how this scales with GMC and galaxy 
properties, as a means to directly constrain the efficiency of feedback as a function of 
these parameters. 

\vspace{-0.7cm}
\acknowledgments 
We thank the anonymous referee for a number of helpful suggestions. Support for PFH was provided by NASA through Einstein Postdoctoral Fellowship Award Number PF1-120083 issued by the Chandra X-ray Observatory Center, which is operated by the Smithsonian Astrophysical Observatory for and on behalf of the NASA under contract NAS8-03060. DN acknowledges partial support from the NSF via grant AST-1009452. EQ is supported in part by NASA grant NNG06GI68G and the David and Lucile Packard Foundation.  \\

\bibliography{/Users/phopkins/Documents/work/papers/ms}

\begin{appendix}
\vspace{-0.5cm}
\section{Resolution and Numerical Tests}
\label{sec:appendix}

Here we briefly discuss some numerical tests of the results in the manuscript. 

In \paperone\ and \papertwo, we consider extensive resolution tests, and demonstrate that global quantities such as the galaxy star formation rate, and correspondingly total IR luminosity, as well as the closely related total mass at ``GMC densities'' ($n\gtrsim100\,{\rm cm^{-3}}$) and by extension CO(1-0) luminosities discussed in the text are well-converged even if we downgrade our numerical resolution by an order of magnitude (see \paperone, Figs.~5, 9, 10, \&\ 11; \papertwo, Appendix~B \&\ Fig.~B1; as well as \citealt{hopkins:stellar.fb.winds}, Appendix~A \&\ Fig.~A3). In these studies we have surveyed our range of galaxy models with independently varied spatial and mass resolution from an order of magnitude more poor than the standard parameters in this text, up to models with $>3\times10^{8}$ particles and sub-parsec resolution. 

However, in this paper we also focus specifically on gas at much higher densities $\sim 10^{4}-10^{5}\,{\rm cm^{-3}}$, which is more challenging to resolve. In {\small GADGET}, the density around a given particle is specifically related to the SPH smoothing kernel size $h_{\rm sml}$ as $\rho_{i} = N_{\rm NGB}\,m_{i}/(4\pi/3\,h_{{\rm sml},\,i}^{3}$), where $m_{i}$ is the particle mass and $N_{\rm NGB}$ the number of SPH ``neighbors.'' Thus resolving a density $n = n_{4}\times10^{4}\,{\rm cm^{-3}}$ requires minimum softening lengths $h_{\rm sml} \lesssim 8.5\,n_{4}^{-1/3}\,(m_{i}/10^{4}\,\msun)^{1/3}\,{\rm pc}$ (with 64 neighbors). This condition is satisfied, although in some cases, only by a relatively small margin, in all our runs at the critical densities used in the text. However, for such resolution to be meaningful, it also must be the case that the simulations can resolve the (turbulent) Jeans mass of structures with these densities (spurious collapse when this is not resolved is prevented by the standard \citealt{truelove:1997.jeans.condition} criterion; see \paperone\ for details). It is straightforward then to see that we are only just able to resolve these high densities in our fiducial runs.

\begin{figure}
    \centering
    \plotonesize{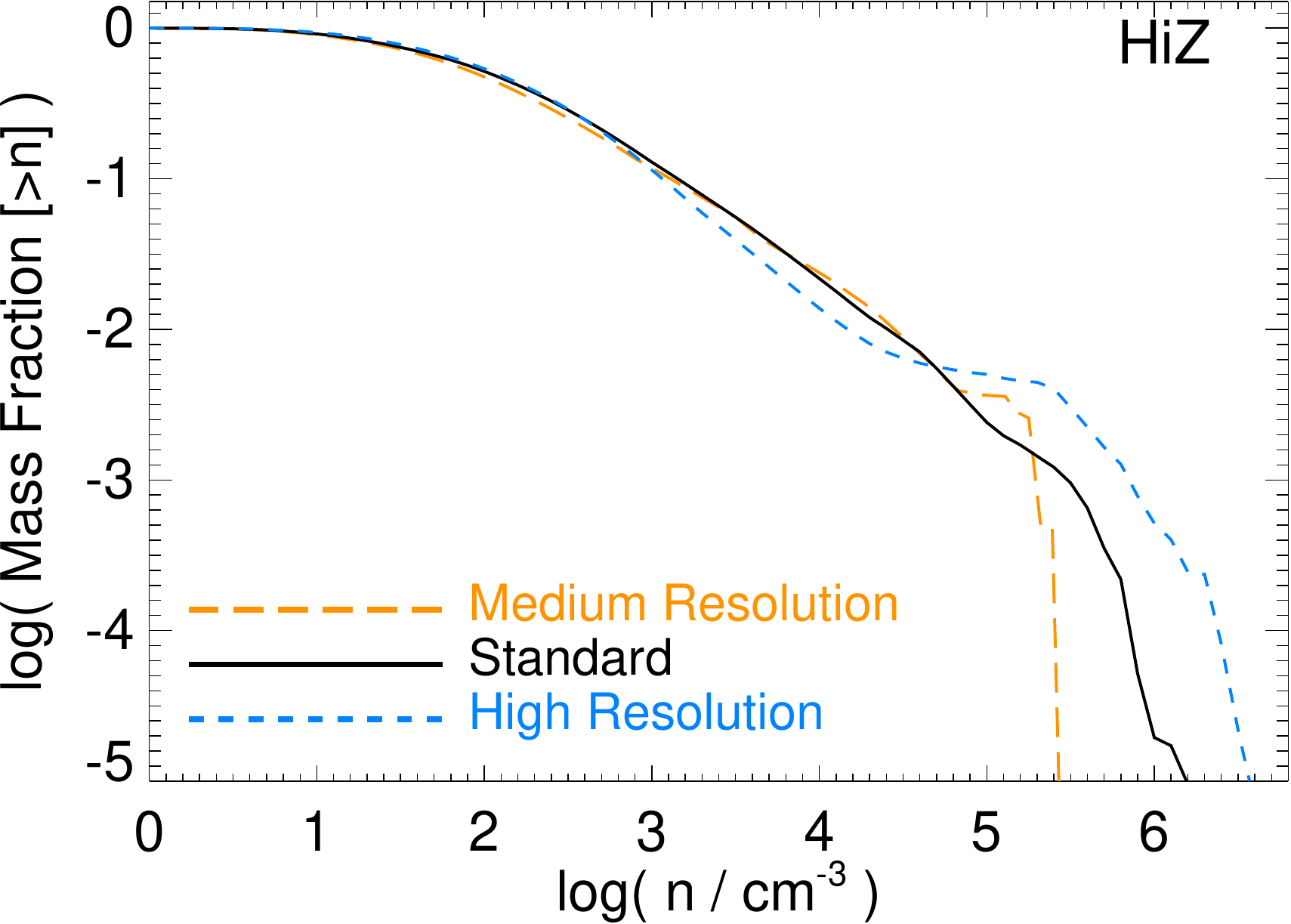}{0.9}
    \caption{Cumulative density distributions from Fig.~\ref{fig:rho.dist.all}, in a resolution study of otherwise identical HiZ models. All examples are our default standard model (with all feedback mechanisms included). The ``standard resolution'' case is run at the same resolution as the model in the main text (for the HiZ model, this is $5\times10^{6}$ gas particles with a force softening of $7\,$pc). We compare ``medium resolution'' ($1.2\times10^{6}$ gas particles, $11\,$pc force softening) and ``high resolution'' ($3\times10^{7}$ gas particles, $4\,$pc force softening). Medium/high resolution runs have also been run for the SMC and MW models, with the same gas particle numbers as those above, but force softenings $=0.2$ and $=0.7$ times smaller, respectively, with qualitatively similar results. 
    \label{fig:rho.dist.res}}
\end{figure}

Fig.~\ref{fig:rho.dist.res} demonstrates this more explicitly by considering a the gas density distribution in a series of HiZ runs with otherwise identical parameters but differing numerical resolution (a run both lower and higher-resolution than our ``standard'' resolution in the text). In each case, the mass and force resolution of the simulation effective sets a maximum resolvable density. At our fiducial resolution this maximum is a few $10^{5}\,{\rm cm^{-3}}$. If we increase our resolution in mass by almost an order of magnitude, this increases to $\sim10^{6}\,{\rm cm^{-3}}$, and if we downgrade it by a similar amount, the density distribution cuts off at $\approx10^{5}\,{\rm cm^{-3}}$. It does appear that at the critical densities $<10^{5}\,{\rm cm^{-3}}$, the results are converged (to within a margin substantially smaller than the scatter in time seen in the simulations). If we use the ``effective densities'' of the high-density tracers (which are somewhat lower than the critical densities), the convergence is better. Clearly, however, going to even higher-density tracers with critical densities $\gtrsim 10^{6}\,{\rm cm^{-3}}$ will require substantially higher-resolution simulations. 

We have also checked whether or not the details of the numerical method change our conclusions. We have re-run a subset of our simulations with the new ``pressure-entropy'' formulation of SPH developed in \citet{hopkins:lagrangian.pressure.sph}. This is an alternative formulation of the SPH equations which removes the spurious numerical ``surface tension'' term and greatly improves the behavior in treating certain fluid mixing instabilities \citep[see also][]{saitoh:2012.dens.indep.sph}. This also eliminates most of the known discrepancies between the results of SPH and Eulerian grid-based simulations. As well, the new version of the code includes improvements in the smoothing kernel \citep{dehnen.aly:2012.sph.kernels}, treatment of artificial viscosity \citep{cullen:2010.inviscid.sph}, and timestepping algorithm \citep{durier:2012.timestep.limiter}. As shown in \citet{hopkins:lagrangian.pressure.sph}, however, these improvements make very little difference to the quantities of interest here (they mostly affect the mixing of diffuse, hot gas in galactic halos); the SFR and dense gas distributions in the galaxies we simulate here are generally altered at the $<20\%$ level.

\end{appendix}

\end{document}